# Image Retrieval Techniques based on Image Features: A State of Art approach for **CBIR**

[1]Mr. Kondekar V. H., [2]Mr. Kolkure V. S., [3]Prof. Kore S.N.
[1]Department of Electronics & Telecommunication Engineering,
([1]Walchand Institute of Technology, Solapur,
[2]Bharat Ratna Indira Gandhi Collage of Engineering, Solapur) Solapur University.
[3]Walchand College of Engineering, Sangali. Shivaji University.

*Abstract-The purpose of this Paper is to describe our research on different feature extraction and matching techniques in designing a Content Based Image Retrieval (CBIR) system. Due to the enormous increase in image database sizes, as well as its vast deployment in various applications, the need for CBIR development arose. Firstly, this paper outlines a description of the primitive feature extraction techniques like: texture, colour, and shape. Once these features are extracted and used as the basis for a similarity check between images, the various matching techniques are discussed. Furthermore, the results of its performance are illustrated by a detailed example.*

## I. INTRODUCTION

As processors become increasingly powerful, and memories become increasingly cheaper, the deployment of large image databases for a variety of applications have now become realisable. Databases of art works, satellite and medical imagery have been attracting more and more users in various professional fields — for example, geography, medicine, architecture, advertising, design, fashion, and publishing. Effectively and efficiently accessing desired images from large and varied image databases is now a necessity

CBIR or Content Based Image Retrieval is the retrieval of images based on visual features such as colour, texture and shape. Reasons for its development are that in many large image databases, traditional methods of image indexing have proven to be insufficient, laborious, and extremely time consuming. These old methods of image indexing, ranging from storing an image in the database and associating it with a keyword or number, to associating it with a categorized description, have become obsolete. This is not CBIR. In CBIR, each image that is stored in the database has its features extracted and compared to the features of the query image. It involves two steps:

- Feature Extraction: The first step in the process is extracting image features to a distinguishable extent.
- Matching: The second step involves matching these features to yield a result that is visually similar.

*Examples of CBIR applications are:*

- Security Check: Finger print or retina scanning for access privileges.
- Intellectual Property: Trademark image registration, where a new candidate mark is compared with existing marks to ensure no risk of confusing property ownership.
- Medical Diagnosis: Using CBIR in a medical database of medical images to aid diagnosis by identifying similar past cases.
- Crime prevention: Automatic face recognition systems, used by police forces.

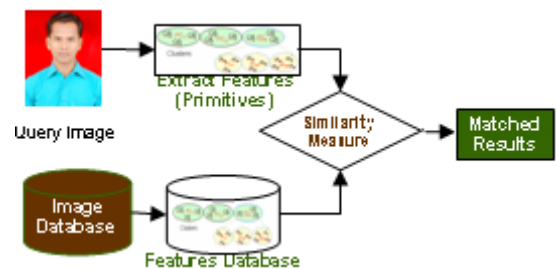

*Fig. 1  CBIR (Content Based Image retrieval ):Block Diagram.*

**Problem Statement**

The problem involves entering an image as a query into a software application that is designed to employ CBIR techniques in extracting visual properties, and matching them. This is done to retrieve images in the database that are visually similar to the query image.

## II. Feature Extraction Techniques.

### 2.1. Colour
#### 2.1.1 Definition

One of the most important features that make possible the recognition of images by humans is colour. Colour is a property that depends on the reflection of light to the eye and the processing of that information in the brain. We use colour everyday to tell the difference between objects, places, and the time of day . Usually colours are defined in three dimensional colour spaces.

These could either be RGB (Red, Green, and Blue), HSV (Hue, Saturation, and Value) or HSB (Hue, Saturation, and Brightness). The last two are dependent on the human perception of hue, saturation, and brightness.

Most image formats such as JPEG, BMP, GIF, use the RGB colour space to store information. The RGB colour space is defined as a unit cube with red, green, and blue axes. Thus, a vector with three co-ordinates represents the colour in this space. When all three coordinates are set to zero the colour perceived is black. When all three coordinates are set to 1 the colour perceived is white. The other colour spaces operate in a similar fashion but with a different perception.

### 2.1.2 Methods of Representation

The main method of representing colour information of images in CBIR systems is through colour histograms. A colour histogram is a type of bar graph, where each bar represents a particular colour of the colour space being used. In MatLab for example you can get a colour histogram of an image in the RGB or HSV colour space. The bars in a colour histogram are referred to as bins and they represent the x-axis. The number of bins depends on the number of colours there are in an image. The y-axis denotes the number of pixels there are in each bin. In other words how many pixels in an image are of a particular colour.

An example of a colour histogram in the HSV colour space can be seen with the image shown. To view a histogram numerically one has to look at the colour map or the numeric representation of each bin. As one can see from the colour map each row represents the colour of a bin. The row is composed of the three coordinates of the colour space. The first coordinate represents hue, the second saturation, and the third, value, thereby giving HSV. The percentages of each of these coordinates are what make up the colour of a bin. Also one can see the corresponding pixel numbers for each bin, which are denoted by the blue lines in the histogram.

Quantization in terms of colour histograms refers to the process of reducing the number of bins by taking colours that are very similar to each other and putting them in the same bin. By default the maximum number of bins one can obtain using the histogram function in MatLab is 256. For the purpose of saving time when trying to compare colour histograms, one can quantize the number of bins. Obviously quantization reduces the information regarding the content of images but as was mentioned this is the trade-off when one wants to reduce processing time.

There are two types of colour histograms, Global colour histograms (GCHs) and Local colour histograms (LCHs). A GCH represents one whole image with a single colour histogram. An LCH divides an image into fixed blocks and takes the colour histogram of each of those blocks. LCHs contain more information about an image but are computationally expensive when comparing images. "The GCH is the traditional method for colour based image retrieval. However, it does not include information concerning the colour distribution of the regions" of an image. Thus when comparing GCHs one might not always get a proper result in terms of similarity of images.

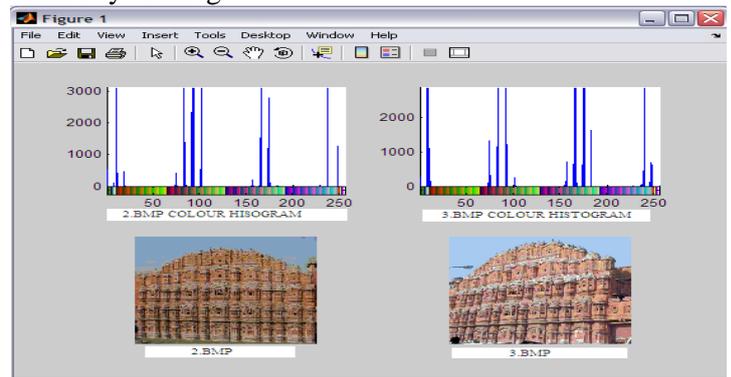
Fig. 2: Histogram of two Image of same class.

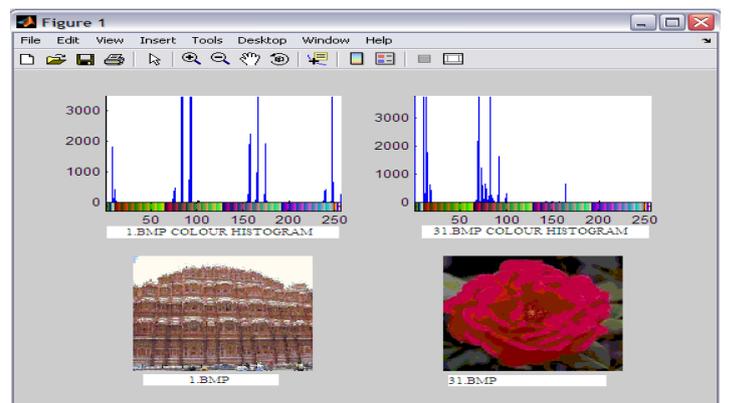
Fig. 3: Histogram of two Images of two different classes.

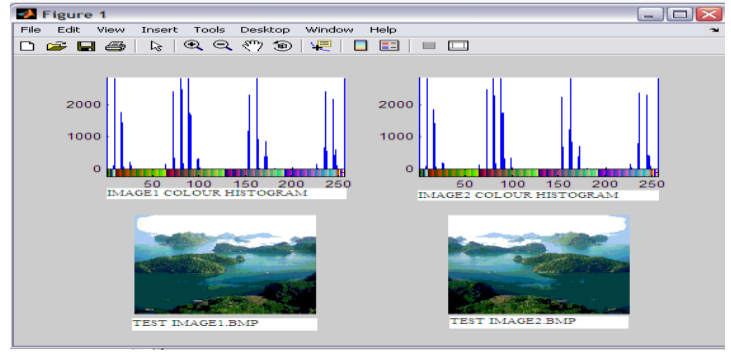
Fig. 4: Histogram of two Image of same class.

## 2. 2 TEXTURE

Texture is an intuitive concept that describes properties like smoothness, coarseness, and regularity of a region. Texture is an important element to human vision, it provides cues to scene depth and surface orientation. In the next sections, Intensity-based texture will be described which has been the topic of investigation for many years and has proven useful. For example, the black and white television proves the usability of Intensity-based texture: people are able to see 3D in a 2D black and white screen. So, it seems important to look at Intensity-based textures before looking at colourful textures because the techniques used by Intensity-based textures can probably be expanded to colour-texture.

### 2.2.1 Texture Definition

Texture is that innate property of all surfaces that describes visual patterns, each having properties of homogeneity. It contains important information about the structural arrangement of the surface, such as; clouds, leaves, bricks, fabric, etc. It also describes the relationship of the surface to the surrounding environment. In short, it is a feature that describes the distinctive physical composition of a surface.

Texture may be defined as a local arrangement of image irradiances projected from a surface patch of perceptually homogeneous irradiances. Texture regions give different interpretations at different distances and at different degrees of visual attention. At a standard distance with normal attention, it gives the notion of macro-regularity that is characteristic of the particular texture. When viewed closely and attentively, homogeneous regions and edges, sometimes constituting texels, are noticeable.

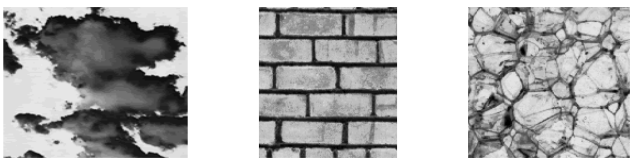

(a) Clouds     (b) Bricks     (c) Rocks

Figure 5: Examples of Texture

### 2.2.3 Texture properties include:

- Coarseness
- Contrast
- Directionality

| Moment | Expression. | Description |
|---|---|---|
| Mean | $m = \sum_{i=0}^{L-1} Z_i P(Z_i)$ | To measure the average intensity |
| Standard deviation | $\sigma = \sqrt{\mu_2(Z)} = \sqrt{\sigma^2}$ | To measure the average contrast |
| Smoothness | $R = 1 - \dfrac{1}{(1+\sigma^2)}$ | To measure the relative smoothness of the intensity in a region. |
| Third moment | $\mu_3 = \sum_{i=0}^{L-1} (Z_i - m)^3 p(Z_i)$ | To measure the skewness of a histogram |
| Uniformity | $U = \sum_{i=0}^{L-1} P^2(Z_i)$ | To measure the uniformity |
| Entropy | $e = -\sum_{i=0}^{L-1} p(Z_i) \log_2 P(Z_i)$ | To measure the randomness |

*Table: 1 statistical parameters*

- Line-likeness
- Regularity
- Roughness

Texture is one of the most important defining features of an image. It is characterized by the spatial distribution of grey levels in a neighbourhood. In order to capture the spatial dependence of gray-level values, which contribute to the perception of texture, a two-dimensional dependence texture analysis matrix is taken into consideration.

This two-dimensional matrix is obtained by decoding the image file; jpeg, bmp, etc.

### 2.2.4 Texture Features:

### Methods of Representation

There are three principal approaches used to describe texture; statistical, structural and spectral.

- Statistical techniques characterize textures using the statistical properties of the grey levels of the points/pixels comprising a surface image. Typically, these properties are computed using: the grey level co-occurrence matrix of the surface, or the wavelet transformation of the surface.
- Structural techniques characterize textures as being composed of simple primitive structures called "texels" (or texture elements). These are arranged regularly on a surface according to some surface arrangement rules.
- Spectral techniques are based on properties of the Fourier spectrum and describe global .periodicity of the grey levels of a surface by Identifying high-energy peaks in the Fourier spectrum.

For optimum classification purposes, what concern are the statistical techniques of characterization. This is because it is these techniques that result in computing texture properties. The most popular statistical representations of texture are:
- Co-occurrence Matrix
- Tamura Texture
- Wavelet Transform.

## 2.3. SHAPE

### 2.3. 1 Definition of Shape:

Shape may be defined as the characteristic surface configuration of an object; an outline or contour. It permits an object to be distinguished from its surroundings by its outline. Shape representations can be generally divided into two categories:
Boundary-based, and Region-based.

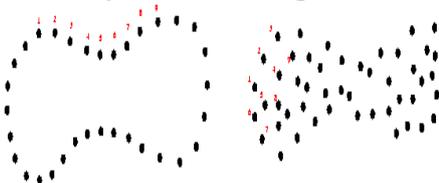

*Figure 6: Boundary-based & Region-based*

Boundary-based shape representation only uses the outer boundary of the shape. This is done by describing the considered region using its external characteristics; i.e., the pixels along the object boundary. Region-based shape representation uses the entire shape region by describing the considered region using its internal characteristics; i.e., the pixels contained in that region.

### 2.3.2 Methods of Representation

For representing shape features mathematically, we have:
Boundary-based:
Polygonal Models, boundary partitioning
Fourier Descriptors:
Splines, higher order constructs
Curvature Models
Region-based:
Super quadrics
Fourier Descriptors
Implicit Polynomials
Blum's skeletons

The most successful representations for shape categories are Fourier Descriptor and Moment Invariants:
The main idea of Fourier Descriptor is to use the Fourier transformed boundary as the shape feature.
The main idea of Moment invariants is to use region-based moments, which are invariant to transformations as the shape feature.

### III. IMAGE FEATURE MATCHING

### 3.1 Similarity Distance measures

CBIR employs low level image features such as color, shape or texture to achieve objective and automatic indexing, in contrast to subjective and manual indexing in traditional image indexing. For contend based image retrieval, the image feature extracted is usually an N-dimensional feature vector which can be regarded as a point in a N-dimensional space. Once images are indexed into the database using the extracted feature vectors, the retrieval of images is essentially the determination of similarity between the features of query image and the features of target images in database, which is essentially the determination of distance between the feature vectors representing the images. The desirable distance measure should reflect human perception. That is to say, perceptually similar images should have smaller distance between them and perceptually different images should have larger distance between them. Therefore, for a given shape feature, the higher the retrieval accuracy, the better the distance measure. Various distance measures have been exploited in image retrieval, are discussed below.

## 3.2 SIMILARITY MEASUREMENTS

A similarity measurement is normally defined as a metric distance. In this section different similarity measurements are described in details.

**Minkowski-form distance metrics**
The Minkowski metric between two point's p = (x1, y1) and q = (x2, y2) is defined as:

$$d^k(P,Q) = (|x_1 - y_1|^K + |x_2 - y_2|^K) \quad (1)$$

**The histogram distance**
The histogram distance, calculated per bin m, between a query image q and a target image t is denoted as:

$$D_i(q,t) = \sum_{m=0}^{M-1} |h_q[m] - h_t[m]| \quad (2)$$

Where M is the total number of bins, hq is the normalized query histogram, and ht is the normalized target histogram. We recognize Di(q; t) as the Minkowski form metric with k=1.

**Histogram the Euclidean distance**
The Euclidean distance is a Minkowski form with k=2:

$$D_e(q,t) = \sqrt{\sum_{m=0}^{M-1}(h_q[m] - h_t[m])^2} \quad (3)$$

The distances (i.e., calculated Minkowski-form distance measures) only take account for the correspondence between each histogram bin. and do not make use of information across bins. This issue has been recognized in histogram matching. As a result, quadratic distance is proposed to take similarity across dimensions into account. It has been reported to provide more desirable result than only matching between similar bins of the color histograms. However, since the histogram quadratic distance computes the cross similarity between colours, it is computationally expensive.

**Histogram quadratic distance**
The quadratic-form distance between two feature vectors q and t is given by:

$$D_e(q,t) = (h_q - h_t)^T A(h_q - h_i) \quad (4)$$

Where A = [aij] is a similarity matrix. aij denotes the similarity between elements with indexes i and j. Please note, that $h_q$ and ht are denoted as vectors.

In order to determine the intersection similarity (S) we adapt Equation to give:

$$s_{q,t} = \sum_{m=0}^{M-1} 1 - |h_q(m) - h_t(m)|. \quad (5)$$

Minkowski-form distance metrics compare only the same bins between color histograms and are defined as:

$$d(Q,I) = \sum_{i=1}^{N} |H_Q[i] - H_I[i]|^r \quad (6)$$

Where Q and I are two images ,N , is the number of bins in the color histogram (for each image we reduce the colours to N , in the RGB color space, so each color histogram has N bins, HQ[i] is the value of bin i in color histogram HQ , which represents the image Q , and HI[i] is the value of bin i in color histogram HI , which represents the image I.

When r=1, the Minkoski-form distance metric becomes L1. When r=2 , the Minkoski-form distance metric becomes the Euclidean distance. In fact, this Euclidean distance can be treated as the spatial distance in a multi-dimensional space.

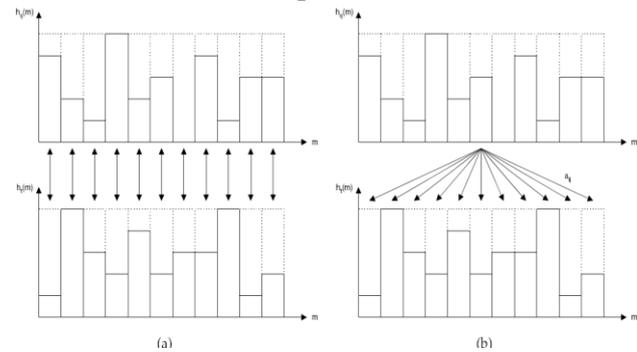

*Figure 7 : (a) Minkowski-form distance metrics compare only similar bins between histograms.*
*(b) Quadratic-form distance metrics compare multiple bins between histograms using similarity matrix A = [aij ].*

**Quadratic-form distance metrics**
The quadratic distance, also called cross distance, is used in the QBIC-system. This method considers the cross-correlation between histogram bins based on the perceptual similarity of the colours represented by the bins. The set of correlation values is represented in a similarity matrix.

Quadratic-form distance metric compares not only the same bins but multiple bins between color histograms and is defined as:

$$d(Q,I) = (H_Q - H_I)^t A(H_Q - H_I) \quad (7)$$

Where Q and I are two images HQ is the color histogram of image Q, HI is the color histogram of image I, |A|=[ai,j] is a NXN 54 , matrix, N is the number of bins in the color histograms, and ai,j denotes the similarity between colours i and j.The similarity matrix is obtained through a complex algorithm:

$$\alpha_{q,i} = 1 - \frac{[(v_q - v_i)^2 + (s_q \cos(h_q) - s_i \cos(h_i))^2 + (s_q \sin(h_q) - s_i \sin(h_i))^2]^{\frac{1}{2}}}{\sqrt{5}}$$

where (hq, sq, vq) and (hi si , vi) represent hue, saturation, and value components for two colours indexed by two histogram bins. Quadratic-form distance metrics overcome a shortcoming of the Minkoski-form distance metrics in that the latter assumes that bins in color histograms are totally unrelated, while the former does not.

**Chebyshev Distance**
This distance calculation metric is named after Pafnuty Lvovich Chebyshev, it is also known as chessboard distance, the equation is defined as

$$d = \max(|x2 - x1|, |y2 - y1|) \quad (8)$$

Bray Curtis distance

$$d_{i,j} = \frac{\sum_{k=1}^{m} |x_{ik} - x_{jk}|}{\sum_{k=1}^{m} |x_{ik} + x_{jk}|} \quad (9)$$

**Manhattan Distance**
Manhattan distance is also known as Taxicab distance. This is because it comes from the fact that it represents the shortest distance a car will drive in a city laid out in square blocks. For example, in the plane, the Manhattan distance between the point P1 with coordinates (x1, y1) and the point P2 at (x2, y2) is

$$|x_1 - x_2| + |y_1 - y_2| \quad (10)$$

*Hamming Distance*

$$HD = \frac{1}{N} \sum_{j=1}^{N} X_j (XOR) Y_j \quad (11)$$

## IV SYSTEM PERFORMANCE EVALUATION.

*Precision and Recall:*

Testing the effectiveness of the image search engine is about testing how well can the search engine retrieve similar images to the query image and how well the system prevents the returned results that are not relevant to the source at all in the user point of view. The big question here is how we know that which image is relevant. Determining whether or not two images are similar is purely up to the user's perception. Human perceptions can easily recognise the similarity between two images although in some cases, different users can give different opinions. Two evaluation measures were used here to evaluate the effectiveness of the image search engine system.

The first measure is Recall. It is a measure of the ability of a system to present all relevant items. The equation for calculating recall is given below:

$$\text{Recall} = \frac{\text{number of relevant items retrieved}}{\text{number of relevant items in collection}} \quad (12)$$

The second measure is Precision. It is a measure of the ability of a system to present only relevant items. The equation for calculating precision is given below.

$$\text{Precision} = \frac{\text{number of relevant items retrieved}}{\text{total number of items retrieved}} \quad (13)$$

*A retrieval score*

A retrieval score can be computed according to the following evaluation criterion: for each query, the system returns the 'x' closest images to the query, including the query image itself (as the distance from the query image to itself is zero). The number of mismatches can be computed as the number of images returned that belong to a class different than that of the query image, in addition to the number of images that belong to the query image class, but that have not been returned by the system. The retrieval score for one class can be then computed as

$$\text{RetrivalScore} = 100 \times \left[1 - \left(\frac{\text{mismatches}}{x}\right)\right]\% \quad (14)$$

## V Conclusions

CBIR at present is still topic of research interest. Different features are used for retrieval of images such as Image colour quadratic distance for image histogram, Image Euclidian distance for image wavelet transform; image Hamming Distance. And corresponding retrieval Recall and Precision parameters are calculated for each feature. For as to increase retrieval efficiency combination of these features should be used instead of using a single feature for image retrieval.

The retrieval efficiency and timing performance can be further increased if the image collection is trained

(pre-processed) and grouped using supervised learning such as classification or unsupervised learning such as clustering. With that, the image with high similarities in the feature space will be group together and result a smaller search space. This will greatly enhance the search time and precision.

**Future work**

*Classification and Clustering:*

The retrieval efficiency and timing performance can be further increased if the image collection is trained (pre-processed) and grouped using supervised learning such as classification or unsupervised learning such as clustering. With that, the image with high similarities in the feature space will be group together and result a smaller search space. This will greatly enhance the search time and precision.

**Acknowledgements**

I would like to take this opportunity to thank my Guide, Prof. S. N. Kore ,Head of electronics Engineering Department Walchand College of Engineering, Sangli for his valuable guidance, constant inspiration, advice, and encouragement throughout the course of this work. I am also thankful to him for his timely and helpful advices to understand the nature of topic and careful reviews on the subject.

Also, I would like to our Principal Dr. S. A. Halkude and Head Dr. S. K. Dixit for their continuous encouragement.

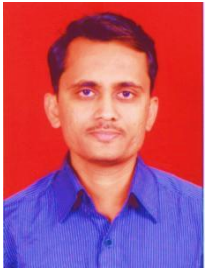
Mr. Kondekar Vipul H. received Master of Engineering in Electronics Engineering with specialization in Computer engineering from Walchand College of Engineering, Sangali Maharastra-India He is working as Lecturer for Last 6 Years in Electronics and Telecommunication Engineering Department at Walchand institute of Technology, Solapur University, Solapur, maharastra-India. His current research interest areas are Image Processing and Microcontroller based system design. He has authored and co-authored more than 10 technical papers published in various prestigious national/international journals and referred conference,symposium,workshop proceedings.

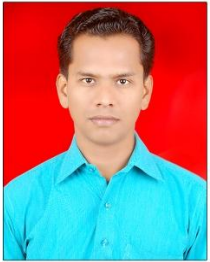
Mr. Kolkure Vijaykumar S. has completed Master of Engineering (ME) in Electronics Engineering with specialization in Computer engineering from Walchand College of Engineering; Sangli Maharastra-India .He has 3 years' experience of working in IT industry. Presently he is working as Lecturer for Last 1 Years in Electronics and Telecommunication Engineering Department at Bharat Ratna Indira Gandhi College of Engineering, Solapur University, Solapur, Maharastra-India. His current research interest areas are Image Processing and Embedded systems. He has more than 8 technical papers published in various prestigious national/international journals and referred conference/symposium/workshop proceedings.